\documentclass[12pt]{iopart} 
\usepackage{amssymb} \usepackage{graphicx} \usepackage{float}
\usepackage{color}

\begin{document}

\title[Nonequilibrium self-organization of colloidal particles on
substrates] {Nonequilibrium self-organization of colloidal particles on
substrates: adsorption, relaxation, and annealing}

\author{Nuno A. M. Ara\'{u}jo, Crist\'{o}v\~{a}o S. Dias, Margarida M.
Telo da Gama}

\address{Centro de F\'isica Te\'orica e Computacional, Faculdade de
Ci\^encias, Universidade de Lisboa, Portugal \\ Departamento de
F\'isica, Faculdade de Ci\^encias, Universidade de Lisboa, Portugal}
\ead{nmaraujo@fc.ul.pt} \vspace{10pt} \begin{indented} \item[]July 2016
\end{indented}

\begin{abstract}
Colloidal particles are considered ideal building blocks to produce
materials with enhanced physical properties. The state-of-the-art
techniques for synthesizing these particles provide control over shape,
size, and directionality of the interactions. In spite of these
advances, there is still a huge gap between the synthesis of individual
components and the management of their spontaneous organization towards
the desired structures. The main challenge is the control over the
dynamics of self-organization. In their kinetic route towards
thermodynamically stable structures, colloidal particles self-organize
into intermediate (mesoscopic) structures that are much larger than the
individual particles and become the relevant units for the dynamics.  To
follow the dynamics and identify kinetically trapped structures, one
needs to develop new theoretical and numerical tools. Here we discuss
the self-organization of functionalized colloids (also known as patchy
colloids) on attractive substrates. We review our recent results on the
adsorption and relaxation and explore the use of annealing cycles to
overcome kinetic barriers and drive the relaxation towards the targeted
structures.
\end{abstract}

%
%
%
%
%

\section{Introduction}
Colloidal particles have a typical size ranging from the micron down to
the nanometer scale.  With state-of-the-art techniques, it is possible
to synthesize them monodisperse in shape and size
\cite{Matijevic2011,Halamek2011}, spherical or asymmetric in shape
\cite{Sacanna2011}, and to fine tune their pairwise interactions
\cite{Yethiraj2003,Duguet2011}.  These impressive experimental advances
have turned colloidal particles into ideal building blocks to produce a
new generation of materials with enhanced physical properties
\cite{Doppelbauer2010,Glotzer2007,Frenkel2011}. At the colloidal scale,
mechanical and optical properties are strongly sensitive to the
particles spatial arrangement. So, there are two main challenges: the
identification of target structures yielding the desired properties and
the design of pathways to synthesize them. By gripping and releasing
colloidal particles with the tip of a microscope some target structures
can be obtained \cite{Kufer2008}, but this is an expensive strategy
limited to very simple configurations and rather small structures
\cite{Parak2011}. Thus, the ultimate goal of Soft Matter is to produce
the desired structures through self-assembly without individual control
of the particles positions \cite{Whitesides2002,Ma2011}. 

The controlled self-organization of colloidal particles into large-scale
structures is still an experimental and theoretical challenge. Most
theoretical studies have considered equilibrium conditions.
Experimentally, this assumption would imply that the characteristic time
of the external perturbations is much larger than the necessary time to
relax towards equilibrium. Such studies benefit from a well-established
theoretical framework and provide valuable information on the system
free energy, self-healing mechanisms, and aging.  However, in practice,
equilibrium is an exception rather than the rule. Real structures are
typically obtained under nonequilibrium conditions and the kinetics of
growth and relaxation cannot be neglected.  In particular, colloidal
particles usually self-organize into mesoscopic kinetic structures much
larger than the individual constituents.  Their relaxation involves
concerted moves and thus, the lifetime of each state increases strongly
with the number of constituents.  When such structures span the entire
system, for example, a gel, the lifetime may diverge with the system
size and the access to the thermodynamic structures is strongly
compromised \cite{Dias2016,Chakrabarti2014,Zaccarelli2006}.  The
probability of observing a given kinetic structure is not necessarily
dependent on its absolute free energy, but mainly on the size and
accessibility of its basin of attraction in the energy manifold. For
very corrugated energy landscapes, the system explores the energy
manifold through thermally activated processes, hoping from basin to
basin, where the mean residence time in each state depends on the
possible paths to leave it and their corresponding energy barriers. New
numerical and theoretical tools are then needed to investigate how
competing timescales take the system away from equilibrium and to
provide insight on the feasibility of the thermodynamic structures.

As an example, we consider the self-organization of colloidal particles
on attractive substrates when relaxation is slower than aggregation, so
the emerging kinetically trapped (percolating) structures prevent the
evolution towards equilibrium. Studies of self-assembly have focused on
the bulk, neglecting boundary effects. Yet, the presence of a surface
improves the control over aggregation
\cite{VanBlaaderen1997,Cadilhe2007,Kinge2008,
Einstein2010,Ganapathy2010,Bernardino2012,Gnan2012,Joshi2016}.  For
example, in the limit of irreversible binding, the assembled structures
differ dramatically from the thermodynamic ones
\cite{Araujo2008,Marques2012,Dias2013,Dias2013a}.  We also consider
colloidal particles with limited valence, known as patchy colloids. The
absence of directional links in isotropic colloids limits the control of
the final structure \cite{Pawar2010,Duguet2011}. By contrast, patchy
colloids, decorated with $n$ attractive patches on their surface, permit
a fine control of colloidal valence and of the local arrangements.  They
are a powerful tool for engineering the assembly of target structures
and give the possibility of obtaining very-low-density networks of
connected particles, which form gels \cite{Zaccarelli2007,Dias2016a}.
In the last years, experimental and theoretical studies have
investigated how the number, type, and distribution of patches can
affect the aggregation process and, consequently, the assembled
structures
\cite{Kretzschmar2011,Sciortino2007,Bianchi2007,Bianchi2008,Russo2009,Ruzicka2011}.
It has been reported that controlling the valence, allows tuning the
density and temperature of both the gas-liquid and sol-gel transitions
\cite{Bianchi2006,Russo2009}. Patchy models with $n$ lower than $12$
have been shown to disfavor dense local configurations, broadening the
region of stability of the liquid phase in the temperature-density plane
\cite{Bianchi2011}. 

The manuscript is organized in the following away. We discuss the limit
of irreversible adsorption in Sec.~\ref{sec::irrev} and the dynamics of
relaxation in Sec.~\ref{sec::dyn}. The possibility of using annealing
cycles to relax kinetically trapped structures is discussed in
Sec.~\ref{sec::ann}. We make some final remarks in
Sec.~\ref{sec::final}.

\section{Irreversible adsorption\label{sec::irrev}}
The dynamics of colloidal particles on substrates results from a balance
between the adsorption/desorption of particles and the
aggregation/relaxation of the colloidal structures. Let us consider
strongly attractive substrates, such that desorption is practically
negligible within the timescale of interest. Thus, the number of
colloidal particles on the substrate increases monotonically in time.
The relevant timescale is the inter-arrival time (inverse flux), which
corresponds to the typical time between two adsorption events and should
depend on the concentration and mobility of the colloidal particles in
solution (bulk). The aggregation/relaxation is a more complex process.
Initially, it depends only on the number and diffusion coefficient of
the particles on the substrate. As the dynamics evolves, aggregates of
different size and shape are formed leading to a hierarchy of relaxation
times. While diffusion of individual particles occurs in the Brownian
timescale (of the order of a second), the relaxation of aggregates might
take hours or even days.  

To study the role of limited valence and the directionality of interactions
on the adsorption of patchy particles on substrates, we have
considered, as a first approximation, the limit of irreversible adsorption
\cite{Lee1994,Mirkin1996,Nykypanchuk2008,DiMichele2014,Wang2012,
Biancaniello2005,Roldan-Vargas2013,Leunissen2011}.  The inter-arrival
time is much shorter than any relaxation timescale and thus the
particles are considered immobile once adsorbed on the substrate. To
address this limit, we have proposed a discrete kinetic model for
aggregation that is a generalization of the well-known ballistic
deposition model to include the directionality of interactions
\cite{Dias2013b}. Accordingly, particles are considered spherical with
$n$ patches equally spaced on their surface. The particle-particle
interaction is pairwise and it is described as an excluded volume
interaction. The patch-patch interaction is short range and
described in a stochastic way, as explained next. We assume that patches
of two different colloids can form bonds only upon collision. We define
an interaction range on the surface of the colloidal particle,
surrounding each patch, parameterized by an angle $\theta$. When two
particles collide, if the contact point is within the interaction range
of their patches, then the two particles form an irreversible bond
mediated by these patches. In the simulation, since only one colloid
moves at a time, the formation of a bond occurs stochastically with
probability $p$, when the moving particle collides within the
interaction range of a patch of the immobile one. $p$ corresponds to the
fraction of the surface of the moving particle covered by the
interaction range of all $n$ patches.  

Particles are sequentially released from a random position above the
growth front and move ballistically (vertically) towards the substrate. If
they hit the substrate without overlapping any previously adsorbed
particle, they immediately stick to it irreversibly with a random
orientation. However, if they collide first with a previously adsorbed
particle, the pairwise interaction described previously is considered.
If the formation of a bond is not successful, the particle is discarded
and a new one is released from a random position above the growth front.  

By neglecting any relaxation, it is possible to access very large length
and timescales. Performing kinetic Monte Carlo simulation, we were able
to show that the kinetic structures are significantly different
from the thermodynamic ones and that they depend strongly on the number
of patches \cite{Dias2013,Dias2015}, patch-patch correlations
\cite{Dias2014,Dias2013a,Dias2014a}, and mechanism of mass transport
\cite{Dias2013b}. We have also identified a very rich critical behavior
of the interfacial properties. In the simplest case of three-patch
particles, we have found two new absorbing phase transitions depending
on the opening angle between the patches~\cite{Dias2014}. A careful scaling
analysis reveals that one of these absorbing phase transitions changes
from discontinuous to continuous at a tricritical flexibility of the bonds
\cite{Araujo2015}. We have also extended our stochastic model to study
particles with distinct patch-patch interactions
\cite{Dias2013a,Dias2014a}. For particles with two types of patches, we
reported a crossover of the interfacial roughness from the
Kardar-Parisi-Zhang (KPZ) to the KPZ with quenched disorder (KPZQ)
universality class when the difference between the strong and weak bonds
is sufficiently large \cite{Dias2014a,Araujo2015}.

\section{Relaxation dynamics\label{sec::dyn}} 
For sufficiently weak particle-particle interactions or long timescales,
the non-equilibrium structures discussed in the previous section are
expected to relax and eventually reach equilibrium. To study such
relaxation dynamics, we consider now a three-dimensional system of
spherical colloidal particles, with three patches equally distributed
along the equator. To follow the individual particles, we performed
Langevin dynamics simulations using LAMMPS \cite{Plimpton1995}. As we
proposed in Ref.~\cite{Dias2016}, colloidal particles are described as
spheres of radius $R$ and mass $m_c$ with three patches fixed at a
distance $R$ from the center, forming an opening angle of $2\pi/3$. The
particles representing the patches are of zero diameter and mass
$10^{-5}m_c$, which is practically negligible for the dynamics. Their
relative position to the center of the core is fixed at all times. The
interaction between colloids is repulsive, described by a Yukawa
potential,
\begin{equation}
V_\mathrm{rep}=
\frac{A}{k}\exp{\left\{-k\left[r-(R_i+R_j)\right]\right\}} ,
\label{eq.Yukawa}
\end{equation}
where $r$ is the distance between the center of the particles, $R_i$ and
$R_j$ are the effective radii of the two interacting particles, $A=1$ is
the interaction strength (in units of $k_BT/2R$) and $k=40$ the inverse
of the screening length (in units of the inverse particle diameter
$(2R)^{-1}$); the interaction between patches is described by an attractive
inverted Gaussian potential~\cite{Dias2016,Vasilyev2013}, defined as,
\begin{equation}
V_\mathrm{att}= -\epsilon\exp\left(-\sigma r_p^2\right) ,
\label{eq.Gauss}
\end{equation}
where $\epsilon=40$ (in units of $k_BT$) is the interaction strength and
$\sigma=10^2$ in units of the inverse squared particle diameter.

The isotropic particle-substrate interaction is obtained from the
Hamaker theory for two spheres~\cite{Everaers2003} in the limit where
the radius of one of the particles diverges. This gives an attractive
$V_\mathrm{s,att}$ and a repulsive $V_\mathrm{s,rep}$ potential, defined
as,
\begin{equation}
 V_\mathrm{s,att}=-\frac{A_H}{6}\left[\frac{2R}{D}\frac{R+D}{D+2R}+\ln\left(\frac{D}{D+2R}\right)\right]
,
\end{equation}
and
\begin{equation}
 V_\mathrm{s,rep}=\frac{A_H\sigma^6}{7560}\left(\frac{6R-D}{D^7}+\frac{D+8R}{(D+2R)^7}\right)
,
\end{equation}
respectively, where $A_H$ is the Hamaker's constant and $D$ the distance
between the surface of the particle and the substrate.

To resolve the stochastic trajectories of the particles, we integrated
the corresponding Langevin equations of motion for the translational and
rotational degrees of freedom,
\begin{eqnarray}
 m\dot{\vec{v}}(t)=-\nabla_{\vec{r}}
V(\vec{r},\vec{\theta})-\frac{m}{\tau_t}\vec{v}(t)+\sqrt{\frac{2mk_BT}{\tau_t}}\vec{\xi}(t),
\\\label{eq.trans_Langevin_dynamics}
 I\dot{\vec{\omega}}(t)=-\nabla_{\vec{\theta}}
V(\vec{r},\vec{\theta})-\frac{I}{\tau_r}\vec{\omega}(t)+\sqrt{\frac{2Ik_BT}{\tau_r}}\vec{\xi}(t),\nonumber
\end{eqnarray}
where, $\vec{v}$ and $\vec{\omega}$ are the translational and angular
velocities, $m$ and $I$ are the mass and inertia of the colloidal
particles, $V$ is the pairwise potential, and $\vec{\xi}(t)$ is the
stochastic term drawn from a random distribution with zero mean. We
considered a relation between damping times $\tau_r=10\tau_t/3$.  For
the sake of generality, time is given in units of the Brownian time,
defined as $\tau_B=(2R)^2/D_t$, where $D_t=k_BT\tau_t/m_c$ is the
translational diffusion coefficient.

\begin{figure}
 \begin{center}
   \includegraphics[width=0.5\textwidth]{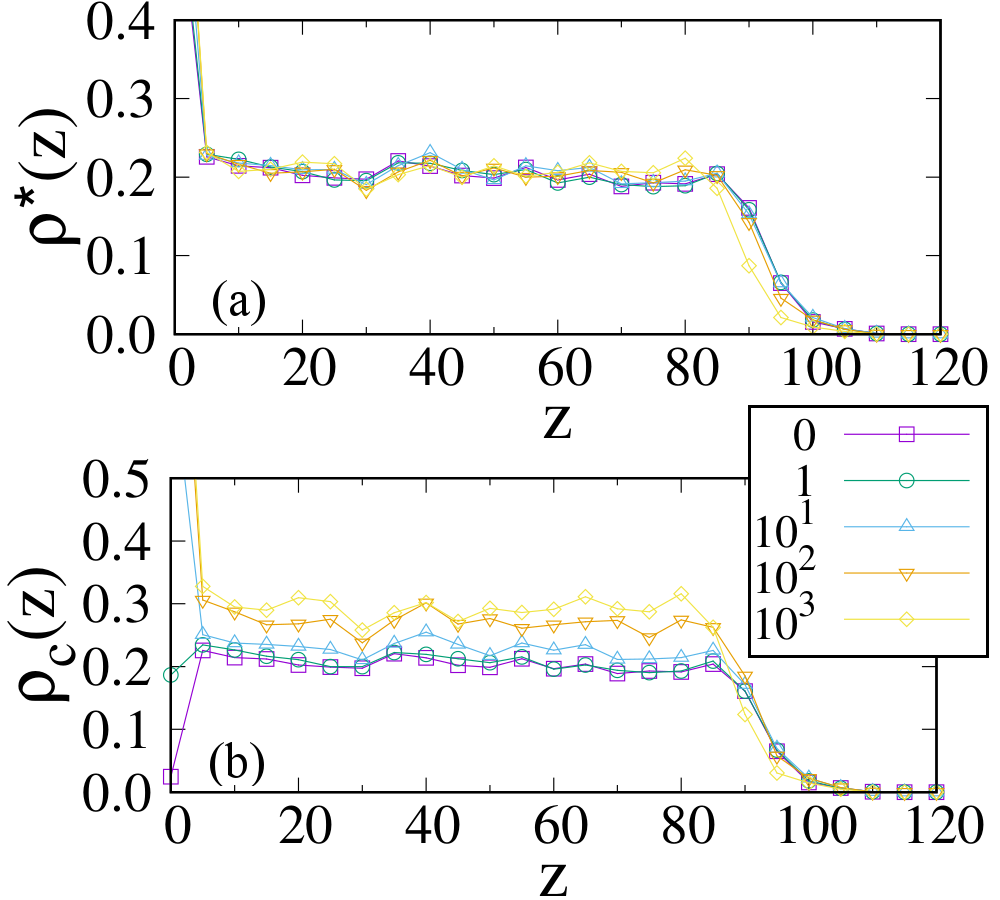} \\
 \end{center}
\caption{(a) Particle and (b) bond density as a function of the distance
to the substrate $z$, at different times, for an initial structure
consisting of $5120$ patchy particles adsorbed on an attractive
substrate, obtained with the stochastic model described in
Sec.~\ref{sec::irrev}. Results are averages over $10$ samples, for
patchy particles adsorbed on a substrate of lateral size $L=16$, in
units of the particle diameter.~\label{fig::density_time}}
\end{figure}
\begin{figure}
 \begin{center}
   \includegraphics[width=0.9\textwidth]{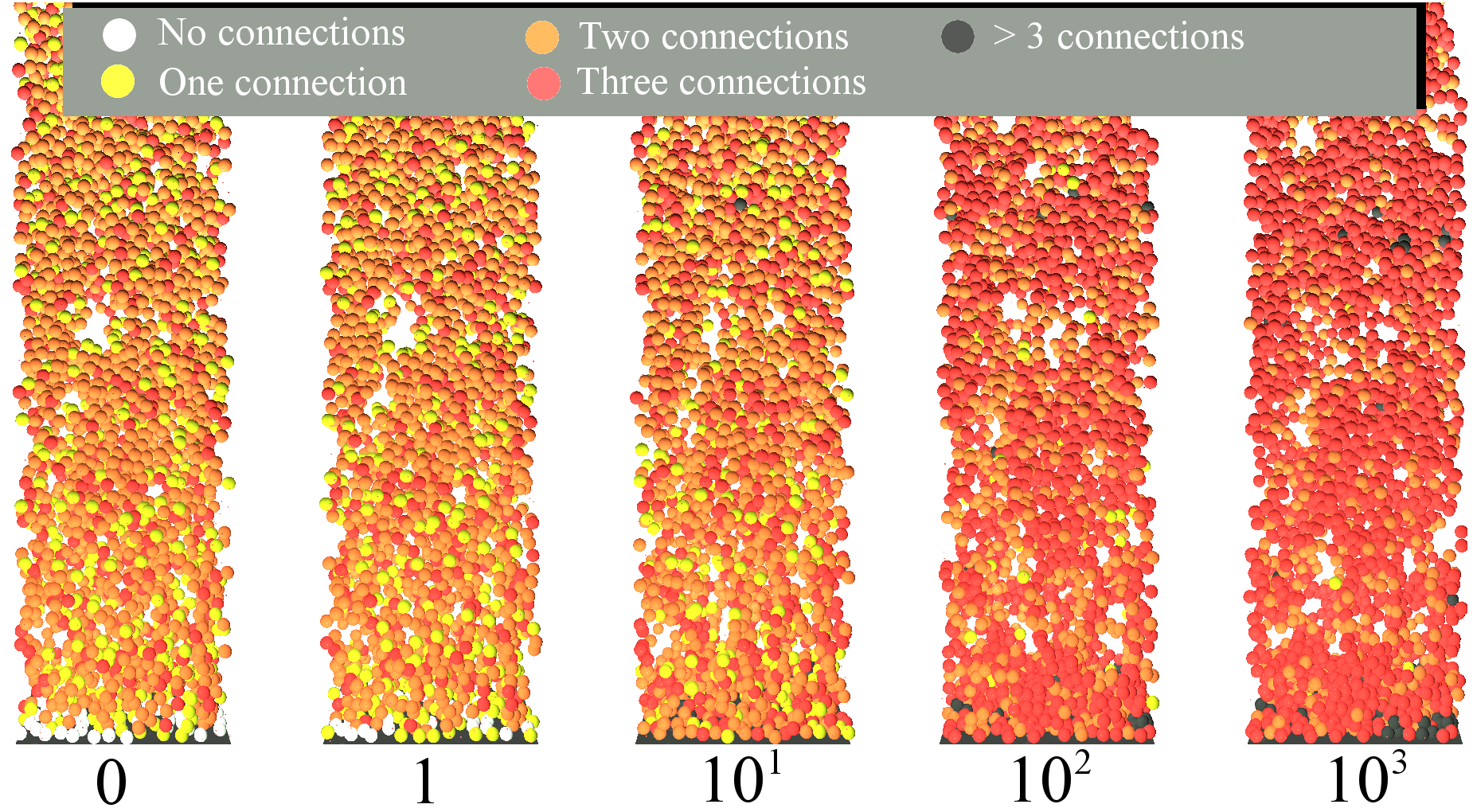} \\
 \end{center}
\caption{Snapshots for $5120$ patchy particles on a substrate of lateral
size $L=16$, in units of the particle diameter. Snapshots are at
different instants of time, namely, $t=\{0,1,10,10^2,10^3\}$, in units
of the Brownian time $\tau_B$. The color of each particle indicates its
number of bonds.~\label{fig::bulk_snaps}}
\end{figure}

The initial structures were obtained with the stochastic model described
in the previous section. They consist of $5120$ particles adsorbed on
a substrate of lateral size $L=16$, in units of the particle diameter. For
the dynamics, we considered $\tau_t= 0.1$ and $k_BT= 1$.
Figure~\ref{fig::density_time} shows, for different instants of time,
(a) the particle density $\rho^*(z)$, defined as the number of particles
per unit volume, and (b) the bond density $\rho_c(z)$, as a function of
the distance to the substrate $z$. While the particle density profile
does not change significantly in time, the bond density does increase by
more than $40\%$. Fig.~\ref{fig::bulk_snaps} depicts snapshots at the
same instants, where the number of bonds per particle is indicated by
the particle color. For the thermostat temperature considered, no bond
breaking is observed within the time scale of the simulation
($10^3\tau_B$). Rather, the relaxation dynamics is driven by the
formation of new bonds, maximizing the number of bonds per particle. As
one starts from a singly connected, tree-like structure, loops are
formed every time a new bond is established, which is expected to affect
the mechanical properties of the colloidal network.

\begin{figure}
 \begin{center}
   \includegraphics[width=0.9\textwidth]{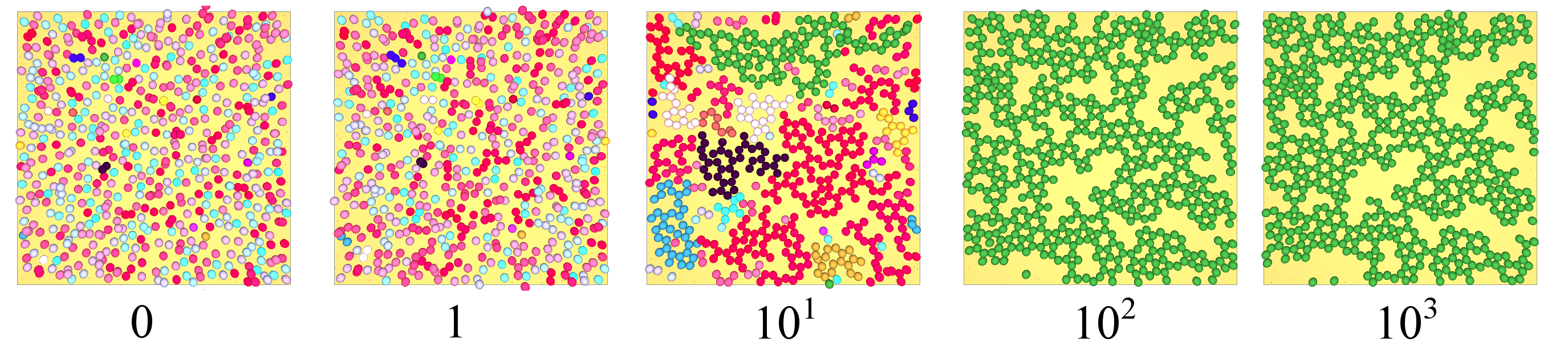} \\
 \end{center}
 \caption{Snapshots for $614$ patchy particles adsorbed on a substrate
of lateral size $L=32$, in units of the particle diameter, at different
instants of time, namely, $t=\{0, 1, 10, 10^2, 10^3\}$, in units of the
Brownian time $\tau_B$. Different colors are different aggregates of
connected particles.~\label{fig::substrate_snaps}}
\end{figure}
\begin{figure}
  \begin{center}
    \includegraphics[width=0.9\textwidth]{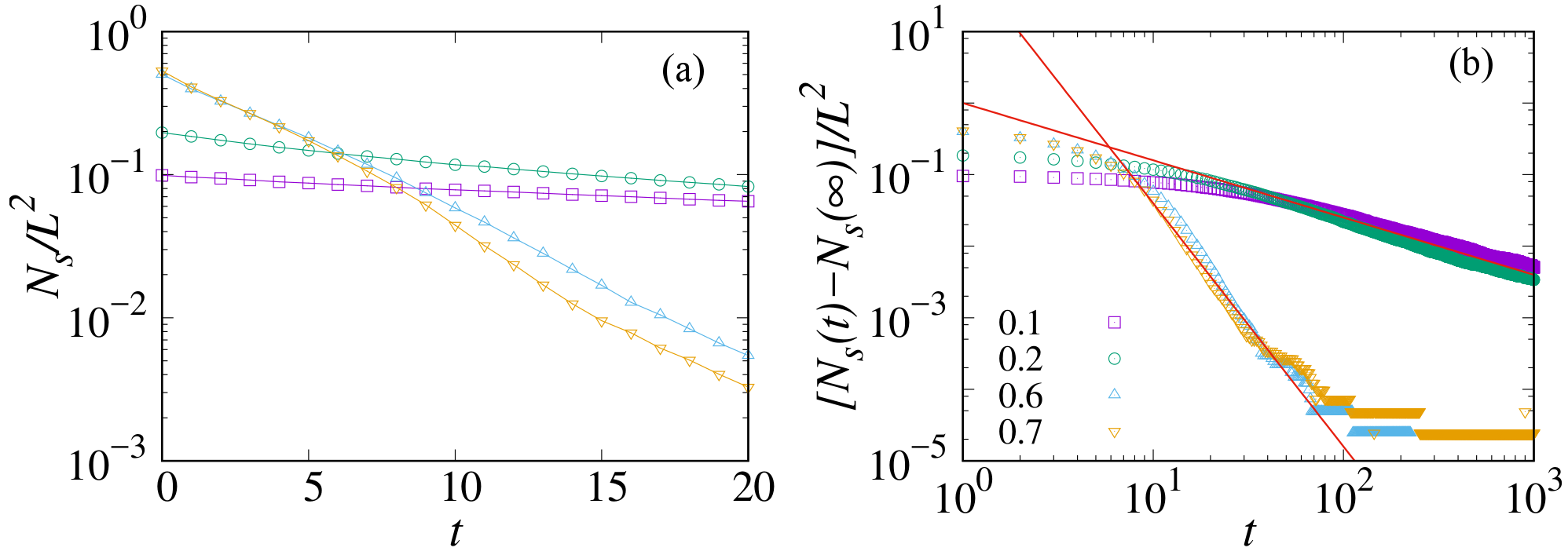} \\
  \end{center}
  \caption{(a) Time evolution of the number of aggregates $N_s$ for
different initial coverages $\theta=N\pi/4$, defined as the ratio of the
total area of the cross-section of all adsorbed particles and the area
of the substrate, with $N=\{0.1, 0.2, 0.6, 0.7\}$. (b) Approach to the
asymptotic value $N_s(\infty)=1$. Results are averages over $10$ samples
on substrates of lateral size $L=64$, in units of the particle diameter.
\label{fig::cluster_relax}}
\end{figure}

To characterize the initial stage of adsorption, in
Ref.~\cite{Dias2016}, we focused on the submonolayer regime, where the
number of adsorbed particles is not enough to completely cover the
substrate. We found that, even in this case, the relaxation towards
thermodynamic phases is hindered by the formation of kinetically trapped
structures that are stable over long timescales (see
Fig.~\ref{fig::substrate_snaps}). From the time evolution of the number
of different aggregates $N_s$, we identified two relaxation regimes.
Initially, $N_s$ decays exponentially with time (see
Fig.~\ref{fig::cluster_relax}(a)), with a characteristic time that
decays linearly with the number of adsorbed particles in the
submonolayer regime. This fast relaxation is essentially driven by the
diffusion and aggregation of initially isolated
particles~\cite{Dias2016}. As the number of such isolated particles
vanishes, the evolution of $N_s(t)$ towards its asymptotic value
$N_s(\infty)=1$ is much slower, as shown in
Fig.~\ref{fig::cluster_relax}(b). For values of the coverage below the
percolation threshold, the numerical data is consistent with a
scale-free relaxation, with an exponent $0.8\pm0.1$, independent of the
value of the coverage. Above the percolation threshold, a single cluster
is rapidly formed, with a structure that is significantly different from
the thermodynamic one.
 
\section{Annealing cycles\label{sec::ann}}
Adsorbed particles diffuse and form patch-patch bonds with other
particles. For the mechanical stability of the final structures, it is
key that these bonds are strong enough; but strong bonding compromises
significantly the relaxation dynamics. For example, DNA mediated bonds
are practically irreversible within the time scale of
interest~\cite{Geerts2010,Wang2012,Leunissen2011,Joshi2016,Dias2016a}.
In this limit, kinetically arrested structures are formed that differ
significantly from the thermodynamic ones~\cite{Dias2016,Dias2016a}.

One promising strategy to overcome these arrested structures is to
design protocols of switching on/off the patch-patch
bonds~\cite{Nykypanchuk2008,Malinge2016,Mirkin1996,Geerts2010,DiMichele2014}. 
For example, DNA mediated bonds are
characterized by a sharp transition in the bonding probability at their
dissociation (or melting) temperature. Thus, bonds can be effectively
switched on/off by performing temperature annealing cycles around the
melting temperature.

\begin{figure}
  \begin{center}
    \includegraphics[width=0.9\textwidth]{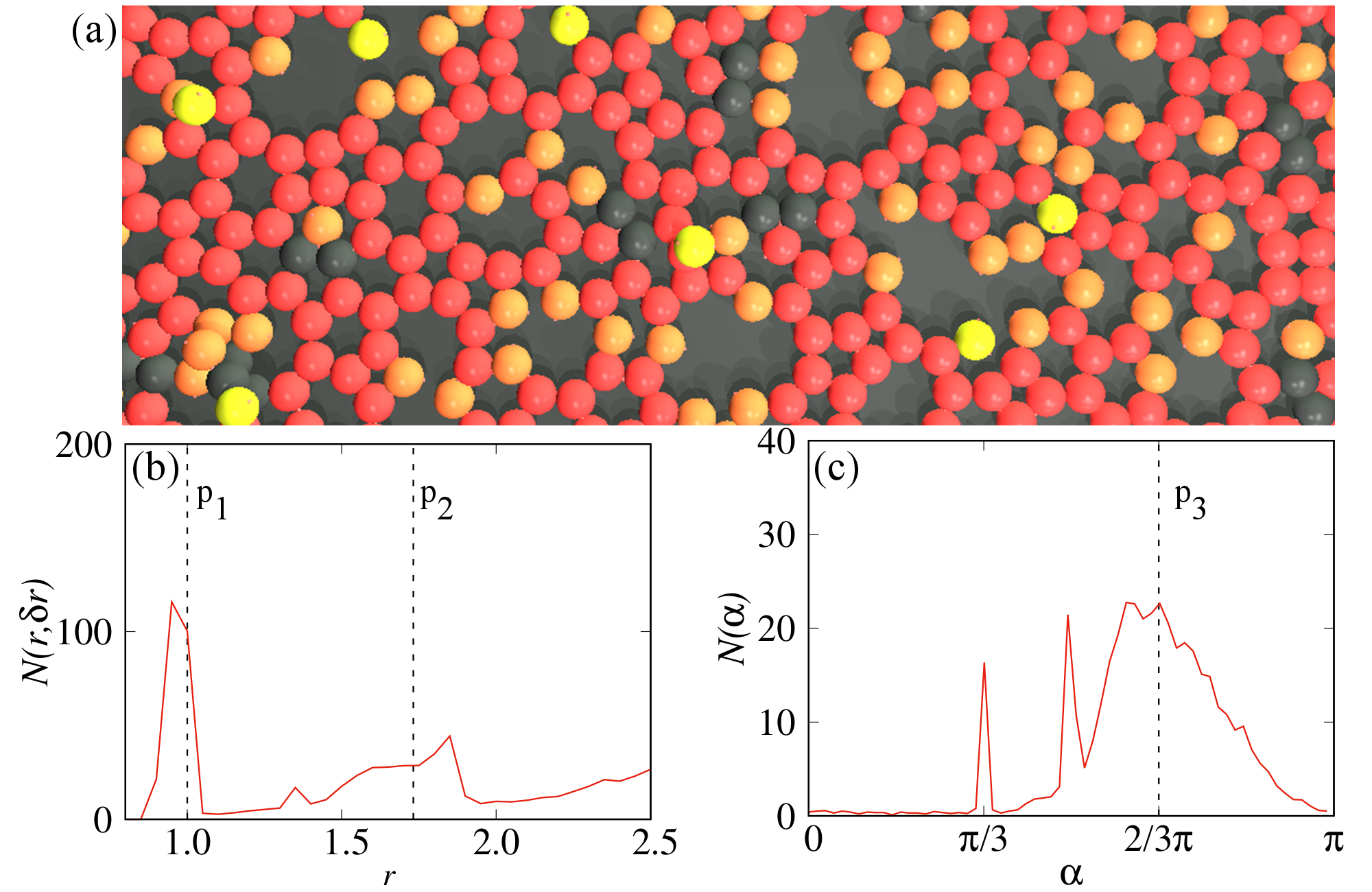} \\
  \end{center}
   \caption{(a) Snapshot, (b) radial and (c) angular distribution
functions for a coverage of $\pi/\left[3\tan\left(\pi/3\right)\right]$,
corresponding to a honeycomb lattice. The vertical dashed lines
correspond to the expected peaks ($p_1, p_2, p_3$) for a honeycomb
lattice: $r=1$, $r=\sqrt{3}$, and $\alpha=2\pi/3$. Results are averages
over $20$ samples on substrates of lateral size $L=32$, in units of the
particle diameter, after $t=10^3$, in units of the Brownian time $\tau_B$,
with $\delta r=0.1$ in
Eq.~(\ref{eq::raddistrib}).~\label{fig::annealing_funcs}}
\end{figure}
To explore this possibility, we consider now the same system,
but start from a number of adsorbed patchy particles that correspond
to the coverage of the honeycomb lattice,
$\pi/\left[3\tan\left(\pi/3\right)\right]$.
Figure~\ref{fig::annealing_funcs}(a) shows a snapshot obtained after
relaxing for $10^3\tau_B$.  Figure~\ref{fig::annealing_funcs}(b) is the
radial distribution functions, defined as,
\begin{equation}\label{eq::raddistrib}
N(r,\delta r)=\sum^{N_p-1}_i\sum_{j>i}^{N_p} g\left(r_{ij}-r,\delta
r\right)/r ,
\end{equation}
where, $N_p$ is the total number of particles, $r_{ij}$ is the distance
between each pair $ij$ of particles and $g\left(r_{ij}-r,\delta
r\right)$ is one, if $\left|r_{ij}-r\right|<\delta r$, and zero,
otherwise. For a honeycomb lattice we expect two peaks: at $r=1$ and
$r=\sqrt{3}$, corresponding to the dashed lines $p_1$ and $p_2$,
respectively. Instead, we observe a much broader distribution.
Figure~\ref{fig::annealing_funcs}(c) is an angular distribution function
for the patch-patch bond orientation, where $\alpha$ is the angle
between two bonds of the same particle. A honeycomb lattice is
characterized by a peak ($p_3$) at $\alpha=2/3\pi$. Clearly, due to
thermal fluctuations and practically irreversible bonds, the obtained
structure differs significantly from the honeycomb lattice.

\begin{figure}
  \begin{center}
    \includegraphics[width=0.9\textwidth]{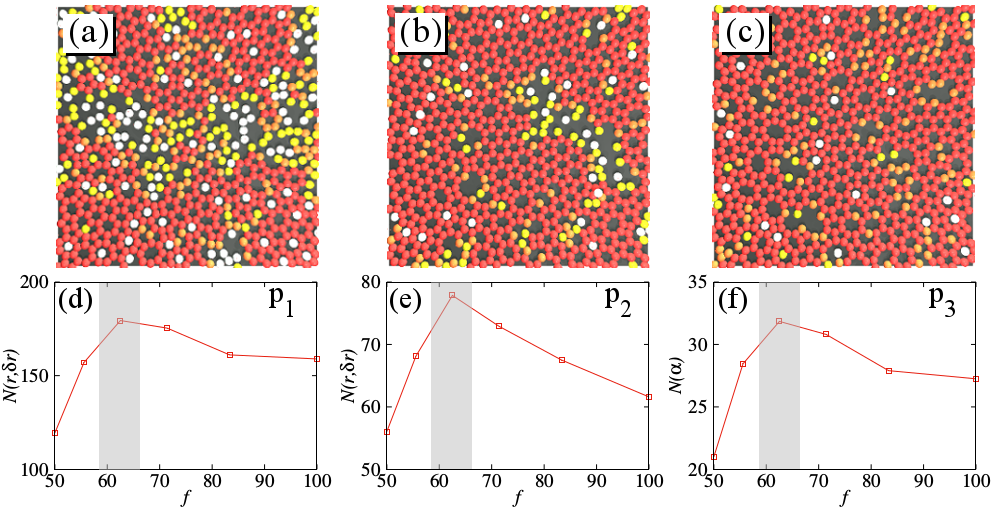} \\
  \end{center}
   \caption{Snapshots for different annealing frequencies $f$, namely,
(a) $50$, (b) $62.5$, and (c) $100$, in units of $\tau_B^{-1}$.  Value
of the radial/angular distribution functions at the peaks (d) $p_1$, (e)
$p_2$, and (f) $p_3$, in Fig.~\ref{fig::annealing_funcs}, as a function
of the annealing frequency $f$. Results are averages over $20$ samples
on substrates of lateral size $L=32$, in units of the particle diameter,
for $t=3000\tau_B$. The system is initially relaxed for $1000\tau_B$,
and the annealing cycles are performed for more $2000\tau_B$ time
units.~\label{fig::anneal_peaks}}
\end{figure} 
To illustrate the advantage of annealing cycles, after an initial
relaxation for $10^3\tau_B$ we performed annealing cycles for an
additional period of $2000\tau_B$. Accordingly, the patch-patch
interaction is switched off for periods of $(2f)^{-1}$, followed by a
period of the same length where the patch-patch interaction is on. Thus,
$f$ is the frequency of the annealing cycle in units of $\tau_B^{-1}$.
Figures~\ref{fig::anneal_peaks}(d)-(e) show the dependence of the peaks
of the radial and angular distribution functions on $f$. We find an
optimal value of $f\approx60$, where the intensity of the peaks is
maximized.

We hypothesize that the optimal frequency results from two competing
mechanisms: The rotational and the translational diffusions. While the
former dominates at higher frequencies; the latter is the dominant one at
lower ones. At high frequencies, the time interval for which the
interaction is off is not long enough for the orientation of two
previously connected patches to decorrelate or two new patches to align.
This alignment is indeed a key mechanism for a particle with only two
bonds to form a third one, since the spatial constraints imposed by the
two connections significantly hinder relaxation. At low frequencies, the
particle-particle correlations decay drastically before bonds are
switched on again and, at every bond-on period of the cycle the dynamics
resembles that at the initial relaxation time.

\section{Final remarks\label{sec::final}}
The dynamics of adsorption and relaxation of patchy particles on
substrates is very rich, involving a hierarchy of processes occurring at
different (length and time) scales. With a stochastic model of
aggregation, it is possible to access the very long scales necessary to
observe critical interfacial phenomena. Studies of the stochastic
(Langevin) dynamics revealed a fast and slow relaxation. The latter
stems from the formation of mesoscopic kinetic structures, the very slow
relaxation of which involves concerted moves, and thus compromises the
feasibility of the thermodynamic structures.  

To overcome these kinetically arrested structures and relax the system
towards the target structures, we have explored the use of annealing
cycles of switching patch-patch interactions on and off. Numerical
results suggest an intermediate optimal frequency at which the target
structural properties are maximized in a given time. Future work should
further explore further this possibility. It is important to
characterize how this optimal frequency depends on the number of
patches, parameters of the particle-particle, patch-patch, and
particle-substrate interaction potentials, and on the number of adsorbed
particles. 

Studies of colloidal particle adsorption on substrates are of scientific
and technological interest. From the theoretical perspective, they are
prototypical examples of collective dynamics under confinement and
require concepts such as competing time and length scales, many-body
correlation, percolation and aging. For practitioners, they provide
valuable information on this feasibility of thermodynamic structures and
on how to design protocols to effectively drive colloidal
self-organization towards the target structures.

\ack
We acknowledge financial support from the Portuguese Foundation for
Science and Technology (FCT) under Contracts nos.
EXCL/FIS-NAN/0083/2012, UID/FIS/00618/2013, and IF/00255/2013.

\section*{References}
\bibliography{SoftMatter}

\providecommand{\newblock}{}
\begin{thebibliography}{10}
\expandafter\ifx\csname url\endcsname\relax
  \def\url#1{{\tt #1}}\fi
\expandafter\ifx\csname urlprefix\endcsname\relax\def\urlprefix{URL }\fi
\providecommand{\eprint}[2][]{\url{#2}}

\bibitem{Matijevic2011}
Matijevic E 2011 {\em {Fine particles in medicine and pharmacy}\/} (New York:
  Springer)

\bibitem{Halamek2011}
Hal{\'{a}}mek J, Zhou J, Hal{\'{a}}mkov{\'{a}} L, Bocharova V, Privman V, Wang
  J and Katz E 2011 {\em Anal. Chem.\/} {\bf 83} 8383

\bibitem{Sacanna2011}
Sacanna S and Pine D~J 2011 {\em Curr. Op. Coll. Interf. Sci.\/} {\bf 16} 96

\bibitem{Yethiraj2003}
Yethiraj A and van Blaaderen A 2003 {\em Nature\/} {\bf 421} 513

\bibitem{Duguet2011}
Duguet E, D{\'{e}}sert A, Perro A and Ravaine S 2011 {\em Chem. Soc. Rev.\/}
  {\bf 40} 941

\bibitem{Doppelbauer2010}
Doppelbauer G, Bianchi E and Kahl G 2010 {\em J. Phys.: Cond. Matt.\/} {\bf 22}
  104105

\bibitem{Glotzer2007}
Glotzer S~C and Solomon M~J 2007 {\em Nat. Mater.\/} {\bf 6} 557

\bibitem{Frenkel2011}
Frenkel D and Wales D~J 2011 {\em Nat. Mater.\/} {\bf 10} 410

\bibitem{Kufer2008}
Kufer S~K, Puchner E~M, Gumpp H, Liedl T and Gaub H~E 2008 {\em Science\/} {\bf
  319} 594

\bibitem{Parak2011}
Parak W~J 2011 {\em Science\/} {\bf 334} 1359

\bibitem{Whitesides2002}
Whitesides G~M and Grzybowski B 2002 {\em Science\/} {\bf 295} 2418

\bibitem{Ma2011}
Ma H and Hao J 2011 {\em Chem. Soc. Rev.\/} {\bf 40} 5457

\bibitem{Dias2016}
Dias C~S, Braga C, Ara{\'{u}}jo N~A~M and {Telo da Gama} M~M 2016 {\em Soft
  Matt.\/} {\bf 12} 1550

\bibitem{Chakrabarti2014}
Chakrabarti D, Kusumaatmaja H, R{\"{u}}hle V and Wales D~J 2014 {\em Phys.
  Chem. Chem. Phys.\/} {\bf 16} 5014

\bibitem{Zaccarelli2006}
Zaccarelli E, Saika-Voivod I, Buldyrev S~V, Moreno A~J, Tartaglia P and
  Sciortino F 2006 {\em J. Chem. Phys.\/} {\bf 124} 124908

\bibitem{VanBlaaderen1997}
van Blaaderen A, Ruel R and Wiltzius P 1997 {\em Nature\/} {\bf 385} 321

\bibitem{Cadilhe2007}
Cadilhe A, Ara{\'{u}}jo N~A~M and Privman V 2007 {\em J. Phys.: Cond. Matter\/}
  {\bf 19} 065124

\bibitem{Kinge2008}
Kinge S, Crego-Calama M and Reinhoudt D~N 2008 {\em ChemPhysChem\/} {\bf 9} 20

\bibitem{Einstein2010}
Einstein T~L and Stasevich T~J 2010 {\em Science\/} {\bf 327} 423

\bibitem{Ganapathy2010}
Ganapathy R, Buckley M~R, Gerbode S~J and Cohen I 2010 {\em Science\/} {\bf
  327} 445

\bibitem{Bernardino2012}
Bernardino N~R and {Telo da Gama} M~M 2012 {\em Phys. Rev. Lett.\/} {\bf 109}
  116103

\bibitem{Gnan2012}
Gnan N, de~las Heras D, Tavares J~M, Telo M~M and Sciortino F 2012 {\em J.
  Chem. Phys.\/} {\bf 084704} 084704

\bibitem{Joshi2016}
Joshi D, Bargteil D, Caciagli A, Burelbach J, Xing Z, Nunes A~S, Pinto D~E~P,
  Ara{\'{u}}jo N~A~M, Bruijc J and Eiser E 2016 {\em Sci. Adv.\/} {\bf 2}
  e1600881

\bibitem{Araujo2008}
Ara{\'{u}}jo N~A~M, Cadilhe A and Privman V 2008 {\em Phys. Rev. E\/} {\bf 77}
  031603

\bibitem{Marques2012}
Marques J~F, Lima A~B, Ara{\'{u}}jo N~A~M and Cadilhe A 2012 {\em Phys. Rev.
  E\/} {\bf 85} 61122

\bibitem{Dias2013}
Dias C~S, Ara{\'{u}}jo N~A~M and {Telo da Gama} M~M 2013 {\em Phys. Rev. E\/}
  {\bf 87} 032308

\bibitem{Dias2013a}
Dias C~S, Ara{\'{u}}jo N~A~M and {Telo da Gama} M~M 2013 {\em Soft Matt.\/}
  {\bf 9} 5616

\bibitem{Pawar2010}
Pawar A~B and Kretzschmar I 2010 {\em Macromol. Rapid Commun.\/} {\bf 31} 150

\bibitem{Zaccarelli2007}
Zaccarelli E 2007 {\em J. Phys.: Condens. Matter\/} {\bf 19} 323101

\bibitem{Dias2016a}
Dias C~S, Tavares J~M, Araujo N~A~M and {Telo da Gama} M~M   arxiv: 1604.05279

\bibitem{Kretzschmar2011}
Kretzschmar I and Song J~H 2011 {\em Curr. Op. Coll. Interf. Sci.\/} {\bf 16}
  84

\bibitem{Sciortino2007}
Sciortino F, Bianchi E, Douglas J~F and Tartaglia P 2007 {\em J. Chem. Phys.\/}
  {\bf 126} 194903

\bibitem{Bianchi2007}
Bianchi E, Tartaglia P, {La Nave} E and Sciortino F 2007 {\em J. Phys. Chem.
  B\/} {\bf 111} 11765

\bibitem{Bianchi2008}
Bianchi E, Tartaglia P, Zaccarelli E and Sciortino F 2008 {\em J. Chem.
  Phys.\/} {\bf 128} 144504

\bibitem{Russo2009}
Russo J, Tartaglia P and Sciortino F 2009 {\em J. Chem. Phys.\/} {\bf 131}
  014504

\bibitem{Ruzicka2011}
Ruzicka B, Zaccarelli E, Zulian L, Angelini R, Sztucki M, Moussa{\"{i}}d A,
  Narayanan T and Sciortino F 2011 {\em Nature Mater.\/} {\bf 10} 56

\bibitem{Bianchi2006}
Bianchi E, Largo J, Tartaglia P, Zaccarelli E and Sciortino F 2006 {\em Phy.
  Rev. Lett.\/} {\bf 97} 168301

\bibitem{Bianchi2011}
Bianchi E, Blaak R and Likos C~N 2011 {\em Phys. Chem. Chem. Phys.\/} {\bf 13}
  6397

\bibitem{Lee1994}
Lee G~U, Chrisey L~A and Colton R~J 1994 {\em Science\/} {\bf 266} 771

\bibitem{Mirkin1996}
Mirkin C~A, Letsinger R~L, Mucic R~C and Storhoff J~J 1996 {\em Nature\/} {\bf
  382} 607

\bibitem{Nykypanchuk2008}
Nykypanchuk D, Maye M~M, van~der Lelie D and Gang O 2008 {\em Nature\/} {\bf
  451} 549

\bibitem{DiMichele2014}
{Di Michele} L, Fiocco D, Varrato F, Sastry S, Eiser E and Foffi G 2014 {\em
  Soft Matt.\/} {\bf 10} 3633

\bibitem{Wang2012}
Wang Y, Breed D~R, Manoharan V~N, Feng L, Hollingsworth A~D, Weck M and Pine
  D~J 2012 {\em Nature\/} {\bf 491} 51

\bibitem{Biancaniello2005}
Biancaniello P~L, Kim A~J and Crocker J~C 2005 {\em Phys. Rev. Lett.\/} {\bf
  94} 058302

\bibitem{Roldan-Vargas2013}
Rold{\'{a}}n-Vargas S, Smallenburg F, Kob W and Sciortino F 2013 {\em Sci.
  Rep.\/} {\bf 3} 2451

\bibitem{Leunissen2011}
Leunissen M~E and Frenkel D 2011 {\em J. Chem. Phys.\/} {\bf 134} 084702

\bibitem{Dias2013b}
Dias C~S, Ara{\'{u}}jo N~A~M and {Telo da Gama} M~M 2013 {\em J. Chem. Phys.\/}
  {\bf 139} 154903

\bibitem{Dias2015}
Dias C~S, Ara{\'{u}}jo N~A~M and {Telo da Gama} M~M 2015 {\em Mol. Phys.\/}
  {\bf 113} 1069

\bibitem{Dias2014}
Dias C~S, Ara{\'{u}}jo N~A~M and {Telo da Gama} M~M 2014 {\em Phys. Rev. E\/}
  {\bf 90} 032302

\bibitem{Dias2014a}
Dias C~S, Ara{\'{u}}jo N~A~M and {Telo da Gama} M~M 2014 {\em Europhys. Lett\/}
  {\bf 107} 56002

\bibitem{Araujo2015}
Ara{\'{u}}jo N~A~M, Dias C~S and {Telo da Gama} M~M 2015 {\em J. Phys.: Cond.
  Matt.\/} {\bf 27} 194123

\bibitem{Plimpton1995}
Plimpton S 1995 {\em J. Comp. Phys.\/} {\bf 117} 1

\bibitem{Vasilyev2013}
Vasilyev O~A, Klumov B~A and Tkachenko A~V 2013 {\em Phys. Rev. E\/} {\bf 88}
  012302

\bibitem{Everaers2003}
Everaers R and Ejtehadi M~R 2003 {\em Phys. Rev. E\/} {\bf 67} 041710

\bibitem{Geerts2010}
Geerts N and Eiser E 2010 {\em Soft Matt.\/} {\bf 6} 4647

\bibitem{Malinge2016}
Malinge J, Mousseau F, Zanchi D, Brun G, Tribet C and Marie E 2016 {\em J.
  Colloid Interf. Sci.\/} {\bf 461} 50

\end{thebibliography}

\end{document}